\documentclass[letter]{aa}  

\usepackage{graphicx}
\usepackage[varg]{txfonts}
\usepackage{appendix}
\usepackage{subfigure}
\usepackage[colorlinks=true,linkcolor=blue,citecolor=blue, urlcolor=blue]{hyperref}
\usepackage[normalem]{ulem}

\begin{document}

   \title{Time to Sparkler}

\subtitle{Accurate ages of lensed globular clusters at $z=1.4$ with JWST photometry}
   \author{Elena Tomasetti
          \inst{1,2}\thanks{e-mail: elena.tomasetti2@unibo.it}
          \and
          Michele Moresco\inst{1,2}
          \and
          Carmela Lardo\inst{1,2}
          \and
          Frédéric Courbin\inst{4,5}
          \and
          Raul Jimenez\inst{4,5}
          \and
          Licia Verde\inst{4,5}
          \and 
          Martin Millon\inst{6, 7}
          \and
          Andrea Cimatti\inst{1,3}}

   \institute{Dipartimento di Fisica e Astronomia “Augusto Righi”–Universit\`a di Bologna, via Piero Gobetti 93/2, I-40129 Bologna, Italy.
         \and
             INAF - Osservatorio di Astrofisica e Scienza dello Spazio di Bologna, via Piero Gobetti 93/3, I-40129 Bologna, Italy.
        \and
            INAF - Osservatorio Astrofisico di Arcetri, Largo E. Fermi 5, I-50125, Firenze, Italy.            
         \and
            ICC, University of Barcelona, Marti i Franques 1, 08028 Barcelona, Spain.
        \and
            ICREA, Pg. Lluis Companys 23, Barcelona, 08010, Spain. 
        \and
        Kavli Institute for Particle Astrophysics and Cosmology and Department of Physics, Stanford University, Stanford, CA 94305, USA.
        \and
        Institute for Particle Physics and Astrophysics, ETH Zurich, Wolfgang-Pauli-Strasse 27, CH-8093 Zurich, Switzerland
             }

   \date{December 9, 2024}

\abstract
  {Determining reliable ages for old stellar objects at different redshifts offers a powerful means to constrain cosmology without relying on a specific cosmological model: this is known as the {\it cosmic clocks} method. Globular clusters, long recognised as hosts of the Universe's oldest stars, have served as the archetypical cosmic clocks. However, their age estimates have traditionally been confined to redshift $z=0$, limiting their role to constraining the present-day age of the Universe $t(z=0)=t_0$. Here we explore how to measure reliable ages of globular clusters well beyond $z = 0$, leveraging their potential to extend cosmic clock measurements to earlier epochs. Specifically, we use 6-band JWST/NIRCam high-precision photometry of candidate stellar clusters in the Sparkler galaxy, located at redshift  $z$ = 1.378 and strongly lensed by the galaxy cluster SMACS J0723.3-7327. By employing stellar population models within a Bayesian inference framework, we constrain the clusters’ ages, star formation histories, metallicities, and dust attenuation. The five compact sources previously identified as globular clusters, based on their red spectral energy distributions being consistent with the colours of old stellar systems, yield a formation age of $1.9 \pm 0.4$ Gyr on average. This result implies a total age of the Universe that aligns well with the $\Lambda$CDM model derived from Planck18 data. Recent space-based observations have uncovered a wealth of lensed globular clusters as well as globulars within the member galaxies of the clusters themselves. These findings suggest that the pool of objects available for cosmic clock studies is enormous. A systematic multi-band photometric survey of globular clusters in and behind galaxy clusters, using facilities like Euclid and the James Webb Space Telescope, would therefore be a powerful tool for estimating cluster ages across a large range of redshifts, allowing the Universe to be dated across an unprecedented range of epochs.}

   \keywords{globular clusters: general – Cosmology: observations – cosmological parameters}

   \maketitle
%

\section{Introduction}

   Galactic globular clusters (GCs) are among the oldest stellar objects in our Galaxy and thus provide a valuable and fully cosmology-independent bound to the age of the Universe which, in turn, can provide stringent constraints on cosmological models \citep[e.g., ][]{Jimenez2019,Valcin2020,Cimatti2023}. 
   GCs are known to be Single Stellar Populations (SSP) in age; while showing evidence for the presence of multiple stellar populations, these remain limited to the chemical composition, especially in light-element abundances\footnote{A small subset of globular clusters, such as $\omega$~Cen, M~54, and Terzan~5, show significant spreads in iron-peak abundances among their stars. These clusters are among the most massive and make up only a few percent of the total globular cluster population \citep[e.g.,][]{Milone2022}.}\citep[e.g.][]{Bastian2018,Gratton2019}.
   Thus the nuclear reactions inside the stars make GCs natural ``nuclear clocks''. The oldest GCs can then be used as {\it cosmic clocks}.
      
   Traditionally, the ages of GCs have been computed from the colour-magnitude diagrams of their resolved stellar populations \citep[e.g.][]{VandenBerg2013,Gallart2005}. This limits their use as cosmic clocks to  $z=0$.
   
  To date, only passively evolving galaxies have provided cosmic clocks at higher redshifts \citep[e.g., ][]{Moresco2012, Moresco2015,Moresco2016,Ratsimbazafy2017,Borghi2022,Tomasetti2023,Jimenez2023}, offering a cosmology-independent constraint on the expansion history of the Universe \citep{Moresco2022}. Recently however, it has been shown that it is possible to obtain precise and accurate ages from GCs integrated light  \citep[e.g.][]{Koleva2008,Cezario2013, Goncalves2020, CabreraZiri2022, Tomasetti2024} which opens up the possibility to measure the ages of extragalactic GCs at $z > 0$ as we will show here.
  Thus, in principle, GCs at high redshift can be cosmic (nuclear) clocks --complementary and independent from passively evolving galaxies-- and yield a robust lower-limit on the age of the Universe at their observed redshift provided their age can be reliably measured.

   GCs are intrinsically faint, with absolute $V$-band magnitudes of $M_{V}=$-7.5 \citep{Baumgardt2020}, they are hardly visible at any redshift $z > 0.1$. At $z=1$ they have apparent magnitudes fainter than $V=$ 32, clearly beyond the reach of any currently planned telescope. Fortunately, one can exploit gravitational lensing as a natural telescope that magnifies the apparent brightness of high-redshift GCs.
   As is the case for high redshift galaxies \citep[Frontier Fields lensing clusters;][]{Lotz2017}, strong gravitational lensing magnifies GC's light, bringing them above the detection limit of current telescopes, in particular JWST.

   Contrary to high-redshift galaxies, the angular extent of GCs is very small -- a typical GC with an effective radius of 50 pc subtends a few milliarcsec at $z=1-1.5$ -- and their stellar population is very uniform across their light profile. This has the double advantage that the magnifications can be very large and there is no chromatic effect due to lensing in the Spectral Energy Distribution (SED) of the GCs. Ages can therefore be measured in a clean way.

   In addition, observing evolved GCs at z$\gg$0.1 complements studies of local GCs \citep[e.g.][]{Tomasetti2024}, with the further advantage that SEDs of younger stellar populations display increased sensitivity to age (see e.g., Fig.~6 in \citealp{Valcin2020}).
    As a result, studying lensed GCs at redshifts $z\sim 1.5$ (corresponding to a third of the current age of the Universe in a vanilla $\Lambda$CDM cosmology) using JWST observations provides a promising solution.

   In the present work, we use for the first time highly magnified GCs in a high-redshift ($z>1$) galaxy as cosmic clocks.
    To do this, we consider the Sparkler galaxy, which is magnified by the $z=0.39$ galaxy cluster SMACS J0723.3-7327 \citep{Pontoppidan2022}. The lensed source, with a magnification of 10--100, is a relatively low-mass galaxy at a redshift of $z = 1.38$ \citep{Caminha2022,Mahler2023}. Its total stellar mass is estimated as $\log (M_*/M_{\odot}) \approx 9.7$, which, after correcting for lensing, translates to an intrinsic mass between $5 \times 10^8$ and  $1 \times 10^9 \, M_{\odot}$ \citep[][for more details]{Mowla2022, Claeyssens2023}. This galaxy has drawn significant attention as it displays prominent compact sources surrounding the main body of the galaxy itself. Most of these compact sources, referred to as ``the sparkles'', remain unresolved even with JWST. Five of them have been identified as GC candidates \citep[e.g.,][]{Mowla2022}, owing to {\em (1)} the absence of [OIII] $\lambda$5007 emission, which is detected in the star-forming regions of the host galaxy but not at the locations of the GC candidates, and {\em (2)} their observed urJ colours, which again suggest that these systems were quiescent at the time of formation. Hereafter we will assume these candidates are indeed GCs.

\begin{figure}[t]
\centering
\includegraphics[width=8.5cm]{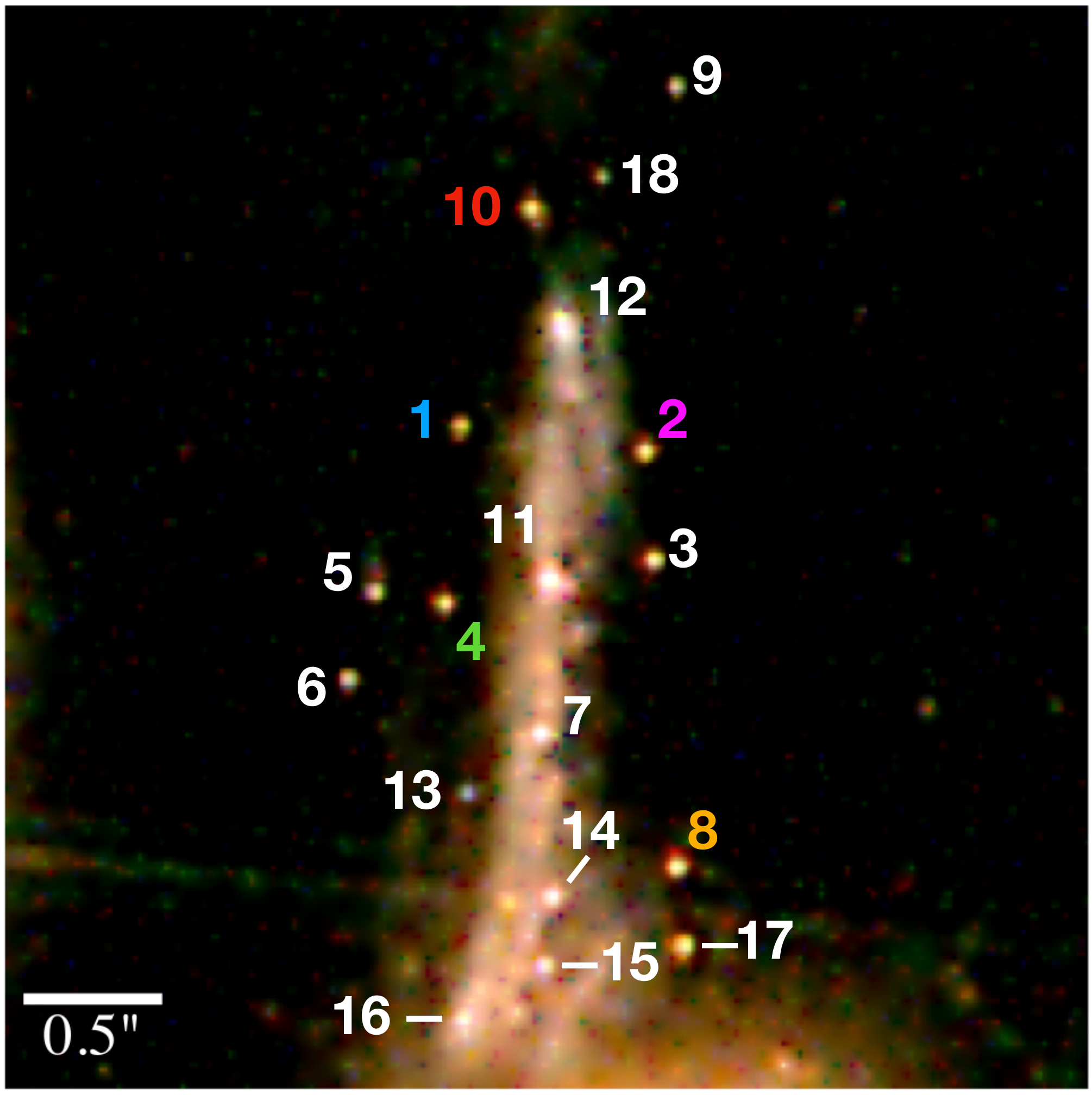}
\caption{Colour composite region of 4\arcsec on-a-side around the Sparkler. The figure is adapted from \citet{Millon2024} and indicates the labeling of the different objects. The colour code corresponds to the one in Fig.~\ref{fig:posteriors}. The PSF in this STARRED-deconvolved image is a circular Gaussian with a FWHM of 0.04\arcsec.}
\label{fig:sparkler}
\end{figure}

Accurate and precise photometry of these objects is crucial to derive reliable ages from SED fitting. However, obtaining reliable photometry of faint point sources superposed on a complex light distribution that changes morphology between bands and in presence of a very complex point-spread function (PSF) is a challenging task.
After the previous analyses of the ``sparkles''  were published \citep{Mowla2022, Claeyssens2023}, STARRED \citep{Millon2024, Michalewicz2023}, a much more sophisticated photometry pipeline, specially developed to handle this challenge, has become available. Here we apply this new methodology to the Sparkler and illustrate how these objects can be used as cosmic clocks.

\section{Data and Methodology}
The data used in this work are the reduced JWST NIRCam images of SMACS J0723.3-7327 in the F090W, F150W, F200W, F277W, F356W and F444W bands provided by~\cite{Mowla2022}. These images have been reduced with a combination of a modified \textit{JWST} pipeline and the {\tt Grizly} software~\citep{grizly} and have a pixel scale of 0.04\arcsec\ per pixel. All the details about image processing are provided in~\cite{Noirot2023}.

\subsection{STARRED photometry}

STARRED \citep{Millon2024, Michalewicz2023} performs deconvolution photometry, making it uniquely well suited to isolate the emission from the sparkles from the lensed arc of the host galaxy. In particular, the STARRED algorithm avoids producing Gibbs oscillations around point sources \citep{Magain1998} and allows the shape of the PSF in the deconvolved image, a.k.a. the {\it target} PSF, to be {\it chosen}.

Because Gaussians do not contain high spatial frequencies, the target PSF is chosen to be a (circular) Gaussian, as opposed to a Dirac function as done in all other algorithms. The deconvolved images are also decomposed into two channels, one containing all analytical Gaussian point sources, and one containing a pixelated image of anything extended. Wavelet regularisation is applied to the latter to enforce the sparsity of the final deconvolved data.

\begin{figure*}[t]
    \centering
    \includegraphics[width=0.99\textwidth]{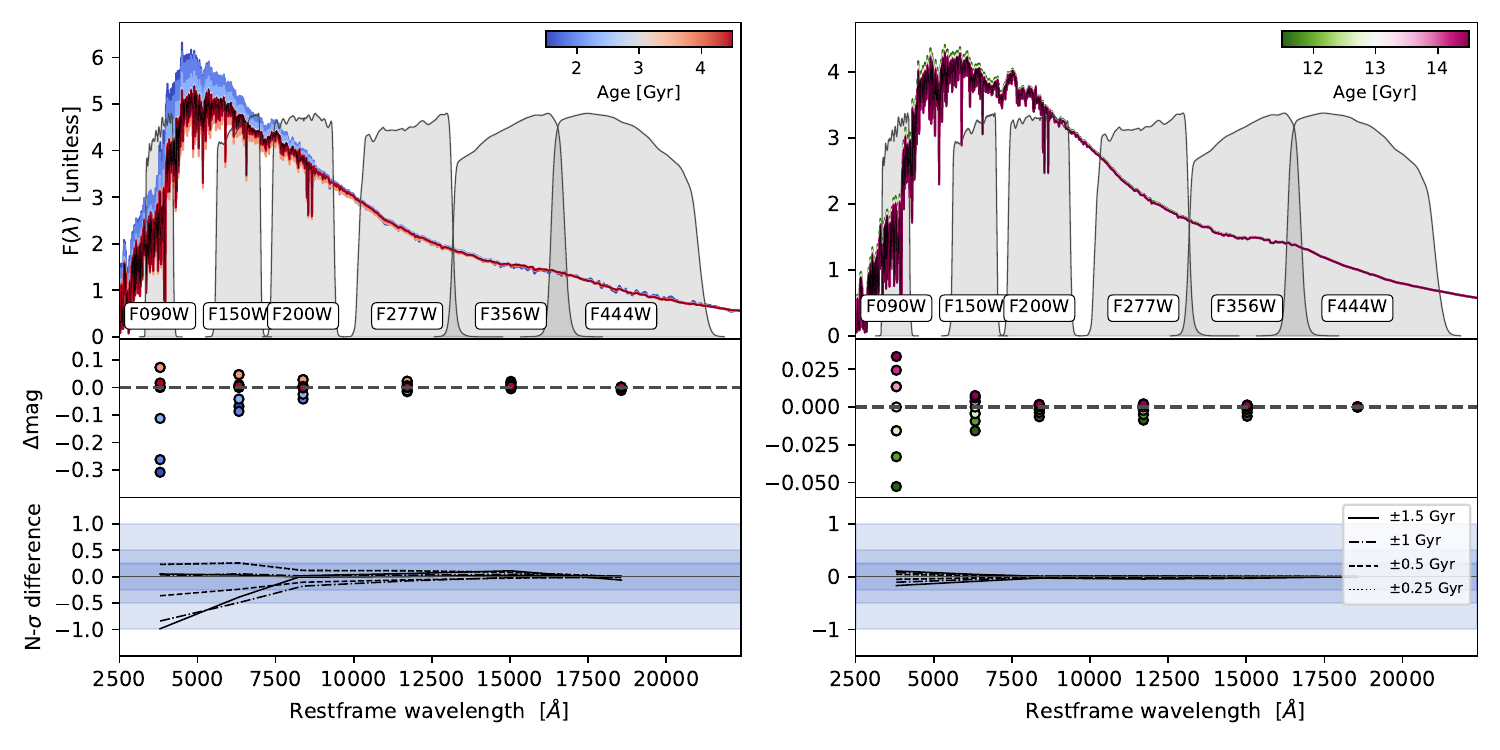} 
    \caption{Sensitivity of GC's SED to absolute age, estimated for the six photometric bands considered in this study. The SEDs are shown in the top panels, for a delayed SFH with $\tau$ = 0.1 Gyr and Kroupa IMF \citep{Kroupa2001} (see Sect. \ref{sec2:fsf} for more details), along with the filter transmission functions used for studying the Sparkler. The left panels show the typical SED of a young population (age $\sim$1.5-4.5 Gyr), while the right panels present an older population (age $\sim$11.5-14.5 Gyr). All SEDs are normalised to the flux in the redder photometric band (the one with the smaller error). The middle panels show the difference in magnitude with respect to a reference age, considering 3 Gyr and 13 Gyr, respectively for the left and right panels. The bottom panels report the significance of the estimated differences normalised by the typical errors in the various bands for the Sparkler's photometry, where the    
    shaded regions show the 0.25, 0.5, and 1$-\sigma$ regions.}
    \label{fig:diff_age}
\end{figure*}

The output is a list of positions and intensities of all point sources along with error bars, and an image of the extended light in the data, with no need to introduce any analytical representation. The same method can be used to obtain very accurate PSFs, even as complex as the JWST ones, including all Airy rings, spikes, and diffraction artifacts. A recent example of an application of this algorithm is light curves of very blended lensed quasars images \citep[see Fig.~2 of][]{Dux2024}. 

The photometry and the PSF used in the present work are the same as in \citep{Millon2024}, which uses the Sparkler as a test case for the method. STARRED was applied to the JWST data for all six bands available for the Sparkler (see Fig.~\ref{fig:sparkler}). 
Note that the spatial resolution achieved on the Sparkler with STARRED is 0.04\arcsec\ (Fig.~\ref{fig:sparkler}). This is at least 10 times larger than the physical size of GCs at $z=1.38$, leaving them undistinguishable from point sources. Indeed, the extended channel of the deconvolved images does not show any trace of residual extended light in any band. 

As STARRED also offers the possibility to compute non-analytical and sub-sampled PSFs as well as a wavelet-regularised treatment of the extended arc, we consider only the STARRED photometry in the following (as listed in Table 3 of \citealp{Millon2024}). In particular, we analyse all 18 compact objects detected by \citet{Millon2024}, including newly identified point sources (IDs 13–18) that are not present in the \citet{Mowla2022} catalogue (see Fig.~\ref{fig:sparkler}). The candidate GCs are identified as Source IDs 1, 2, 4, 8, and 10. Notably, Source IDs 5, 6, 7, 11, and 12 appear extended, with flux residuals observed in the extended channel near these point sources \citep[see][for further discussion]{Millon2024}. While it is not clear whether these spatial extensions are physically associated with the objects or if they are unrelated, STARRED decontaminates by construction the flux of the point-like objects from the flux of anything present in the extended channel of the deconvolved image. 

Finally, let us note that the magnification by the lensing galaxy cluster is about $\mu = 12$ \citep{Claeyssens2023} all over the field of view of Fig.~\ref{fig:sparkler}, illustrating that the lensing caustic is shallow and that chromatic effects due to lensing are negligible, even if intrinsic colour gradients would be present in the lensed sources. 

\subsection{Determining ages and metallicities}\label{sec2:fsf}
\begin{figure*}[t!]
    \centering
    \begin{subfigure}
         \centering
         \includegraphics[width=0.49\textwidth]{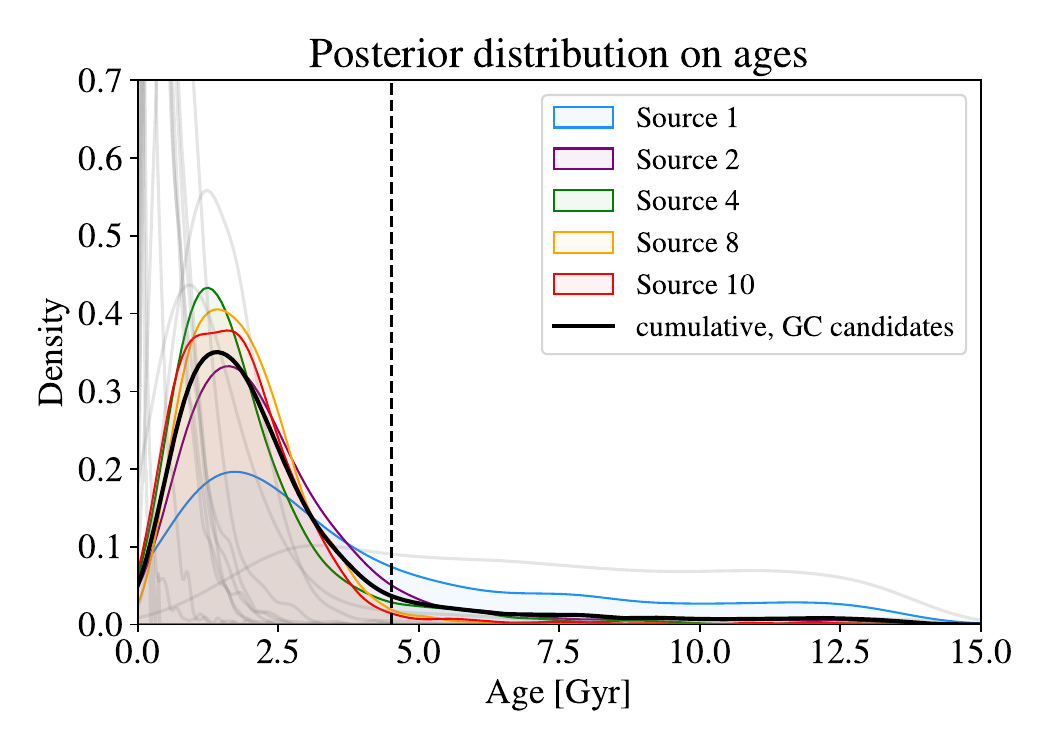} 
         \label{fig:age_pdf}
    \end{subfigure}
    \begin{subfigure}
        \centering
        \includegraphics[width=0.47\textwidth]{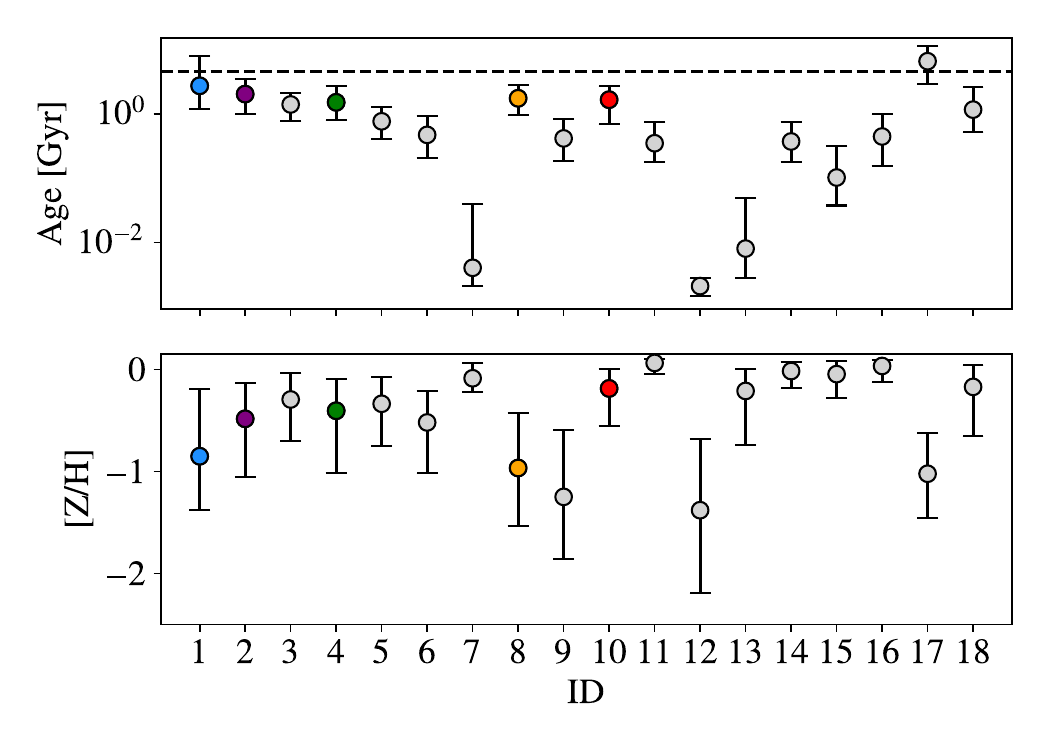}
        \label{fig:age&ZH}
    \end{subfigure}
    \caption{Left panel: Posterior for the ages of the five GCs in the Sparkler (with the posterior distributions for the other sources shown as thick grey lines). The best fit and 68\% credible regions for the ages and metallicities are reported in the right panels. For reference, in both panels, the dashed lines indicate the age of the Universe in a vanilla $\Lambda$CDM cosmology.}
    \label{fig:posteriors}
\end{figure*}
We measure the age, metallicity, and dust reddening of all the sources detected by \citet{Millon2024} using the public code \texttt{BAGPIPES} \citep{Carnall2018}, which allows us to fit photometry adopting a parametric Bayesian approach, maximising the posterior probability via the nested sampling algorithm \texttt{Multinest} \citep{Buchner2016}. A detailed description of all the code's features is presented in \citet{Carnall2019} and \citet{Carnall2022}. 

In this work, the synthetic spectra modeled in \texttt{BAGPIPES} to fit the photometry are based on three main components. \\The first component involves the use of the 2016 version of the \citet{Bruzual&Charlot2003} stellar population models (hereafter BC16, see \citealp{Chevallard2016}), which adopt a \citet{Kroupa2001} initial mass function (IMF). \\The second component is the assumed star formation history (SFH) of the stellar population, which follows a delayed exponentially declining law, $\propto (t-T_0)\exp^{-(t-T_0)/\tau}$. In this model, $T_0$ represents the age of the Universe at the onset of star formation, while $\tau$ sets the width of the SFH. We adopt a uniform prior for $\tau$, ranging from 0 to 1 Gyr. In this work, a key modification to \texttt{BAGPIPES}, already tested and validated in \citet{Jiao2023} and \citet{ Tomasetti2023}, is applied to allow $T_0$, and consequently the age of the stellar population, to span the full range of 0-15 Gyr independently of redshift, thus removing the effects of cosmological assumptions in the age priors.
\\The third component is dust absorption, modeled with a \citet{Calzetti2000} law to account for the potential presence of dust in the system. A uniform prior is adopted on the parameter $A_V$ in the range 0 -- 4 mag.

In total we vary six parameters: 1) the age of the oldest stellar population, 2) the SFH width ($\tau$), 3) overall metallicity [Z/H] = $\log \left(\frac{Z}{Z_\odot}\right)$, 
4) dust attenuation ($\rm{{A_V}}$), 5) velocity dispersion ($\sigma_{\mathrm{v}}$) and 6) stellar mass. 
Stellar mass and velocity dispersion are effectively nuisance parameters that have no effect on the other parameters and therefore are not reported.
Before fitting, we correct the flux in each filter to account for the Milky Way's foreground extinction, listed in Table 1 of \citet{Claeyssens2023}.

It is worth observing here that in the age range typical of objects at z$\sim$1.4 population synthesis models can effectively distinguish between younger and intermediate-age populations, which feature intermediate-mass stars with unique photospheric properties. Notably, it is not just the main-sequence turn-off but also the entire sub-giant branch that is sensitive to age. On the other hand, age determination becomes increasingly challenging for older populations reaching 12 Gyr or more. This is illustrated in Fig.~\ref{fig:diff_age} where the sensitivity of a GC spectrum to a change in age is shown comparing two different populations, a younger ($\sim$3 Gyr) and an older one ($\sim$13 Gyr). All the spectra are normalised to the Sparkler band with the smaller photometric error (F444W). It is evident how the younger populations can constrain a difference in ages of $\pm$0.5-1.5 Gyr with a considerably higher significance than older populations.

\section{Results}
Figure ~\ref{fig:posteriors} summarises the main results.  For all 18 sources, we show the posterior distribution of the ages on the left panel, and on the right panels the central value and the 68.4\% confidence interval for the age and metallicity. GC candidates are highlighted in bright colours. Remarkably, we find that the choice of the upper limit on the prior age does not affect in any significant way the 1-- and 2--$\sigma$ confidence intervals: despite adopting a very wide prior for the ages (uniform between 0 and 15 Gyr) all the recovered ages are compatible with $\Lambda$CDM model's predictions.
For all the parameters except $\tau$ the posteriors are not prior dominated (see appendix for more details). 

The average age of the GC candidates is 1.9 $\pm$ 0.4 Gyr, where the error is the standard deviation. 
For reference, the $\Lambda$CDM Planck18-model inferred age of the Universe at redshift z = 1.378 is $\sim$4.5 Gyr.
The GC candidates show a mean metallicity [Z/H] = -0.6 $\pm$ 0.3. Our derived metallicities are consistent with those measured in \citet{Mowla2022} within the errors, although we find them generally lower by -0.13 $\pm$ 0.27 dex. 
The measured ages and metallicities for the candidate GCs are reported in Table~\ref{tab:results}.

Although beyond the scope of the present paper, the measured ages and metallicities provided here enable the derivation of an age-metallicity relation for the stellar clusters in the Sparkler, making them of broad interest for studies on galaxy formation and chemical enrichment \citep{Forbes2023}.

The derived dust reddening is fairly low for the GC candidates, $\rm{A_V} = 0.3 \pm 0.1$ mag on average. Non-isolated sources, instead, often show reddening values above 0.5 mag (e.g., sources 7, 11, 12, 16, and 17). To assess how the inclusion of dust reddening impacts the age and metallicity of the GC candidates, we also performed the analysis excluding it from the modelling. On average, we find an increment of $\sim$1.1 Gyr in age and $\sim$0.13 dex in metallicity.

\begin{table}
\centering
\renewcommand{\arraystretch}{1.5}
\caption{Results of the SED fitting for the five candidate GCs. The adopted uniform priors are indicated as U(x,y) with x,y the lower and upper limits. The median value and 68.4\% confidence interval for age and metallicity are reported for each source.}
\label{tab:results}
\begin{tabular}{c c  c r r}
\hline
\hline

ID  & Age [Gyr] & [Z/H]\\
\hline
 PRIOR &   U(0,15) & U(-4,0.1)\\
\hline

1  & $2.73^{+5.22}_{-1.53}$ & $-0.85^{+0.66}_{-0.53}$ \\
2  & $2.03^{+1.50}_{-1.04}$ & $-0.48^{+0.35}_{-0.57}$ \\
4  & $1.51^{+1.24}_{-0.71}$ & $-0.40^{+0.31}_{-0.61}$ \\
8  & $1.74^{+1.09}_{-0.78}$ & $-0.96^{+0.54}_{-0.57}$ \\
10 & $1.66^{+1.03}_{-0.97}$ & $-0.18^{+0.19}_{-0.36}$ \\

\hline
\end{tabular}
\end{table}

\section{Conclusions and future prospects}
\label{conclusion}

Cosmic clocks \citep{JimenezLoeb} provide a strictly cosmology-independent constraint on the expansion history of the Universe. While cosmic clocks at $z>1$  were so far exclusively passively evolving galaxies, we have explored the possibility of obtaining absolute ages for strongly lensed and magnified GCs at redshift beyond zero in the lensed Sparkler system, using their integrated six-band photometry from space. The wavelet-based deconvolution-photometry algorithm adopted (STARRED) is uniquely suited to isolate the emission from point-like sources from the complex morphology of the lensed arc of the host galaxy.

For all lensed Sparkler objects, using a fully Bayesian pipeline, we estimated key parameters, such as age, metallicity, and dust attenuation. Very broad priors (especially on the age) were adopted so as not to introduce cosmological biases. The simultaneous fit of the physical parameters of the GCs and the broad priors adopted are key novel aspects of this work. 

GC candidates are point-like and older compared to the other sources. 
Our main result is that the mean age of the five GC candidates is 1.9 $\pm$ 0.4 Gyr, fully consistent with the $\Lambda$CDM Planck18-model prediction for the age of the Universe at redshift $z=1.38$. 
 
This finding opens up the possibility of using GCs at high-$z$ as cosmic clocks to constrain the cosmology model, should more Sparkler-like magnified GC be observed at different redshifts. Interestingly, the uncertainty in the age determination for these GCs today is comparable to that of the first passively evolving galaxies for which age was obtained at a similar redshift $z = 1.5$ in the mid-90s \citep{Dunlop1996}. 
 New JWST/NIRSpec (IFU) spectroscopic observations of the Sparkler have been acquired (Cycle 2 GO\#2969; PI Mowla). These spectra, to which we do not have access at the time of the present study, will represent a strong blind test and validation of our methodology which is based solely on photometry. As photometry is feasible with JWST over large fields of view,  these five GC candidates are the tip of the iceberg of a much bigger population of GCs available for cosmic clock studies. JWST multi-band imaging of galaxy clusters shows a plethora not only of lensed GCs but also of GCs in the member galaxies of the clusters themselves, potentially providing GC samples at redshifts typical for galaxy clusters ($0.1 < z < 0.8$) and for lensed sources well beyond $z=1$. Finally, Euclid images such as the Perseus cluster taken as part of the Early Release Observations data also display numerous (non-lensed) GCs around the very low redshift cluster galaxies \citep{Cuillandre2024}. 

 A systematic multi-band photometric campaign of GCs in and behind galaxy clusters based on Euclid and JWST can enable the measurement of ages of a sizeable population of GCs spanning a broad range of redshifts, which can then be used as cosmic clocks.

\begin{acknowledgements}
ET acknowledges the support from COST Action CA21136 – “Addressing observational tensions in cosmology with systematics and fundamental physics (CosmoVerse)”, supported by COST (European Cooperation in Science and Technology).
Funding for the work of RJ and LV was partially provided by
project PID2022-141125NB-I00, and the “Center of Excellence Maria de Maeztu 2020-2023” award to the
ICCUB (CEX2019- 000918-M) funded by MCIN/AEI/10.13039/501100011033. 
MM acknowledges support from MIUR, PRIN 2022 (grant 2022NY2ZRS 001). MM and AC acknowledge support  from the grant ASI n. 2024-10-HH.0 “Attività scientifiche per la missione Euclid – fase E”.
\end{acknowledgements}

\section*{Appendix}
\begin{figure*}
    \centering
    \begin{subfigure}
         \centering
         \includegraphics[width=0.4\textwidth]{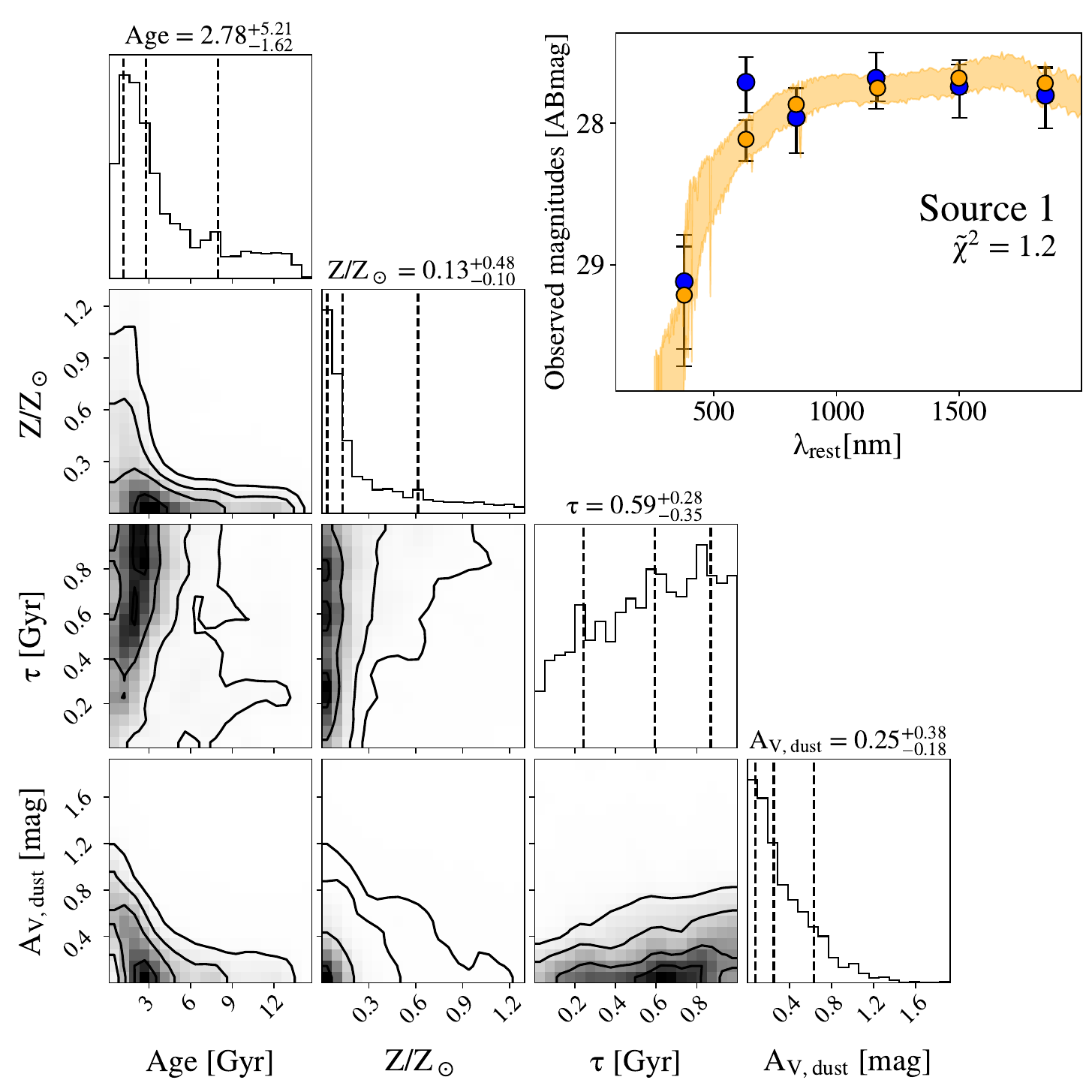} 
         \label{fig:1_corner}
    \end{subfigure}
    \begin{subfigure}
         \centering
         \includegraphics[width=0.4\textwidth]{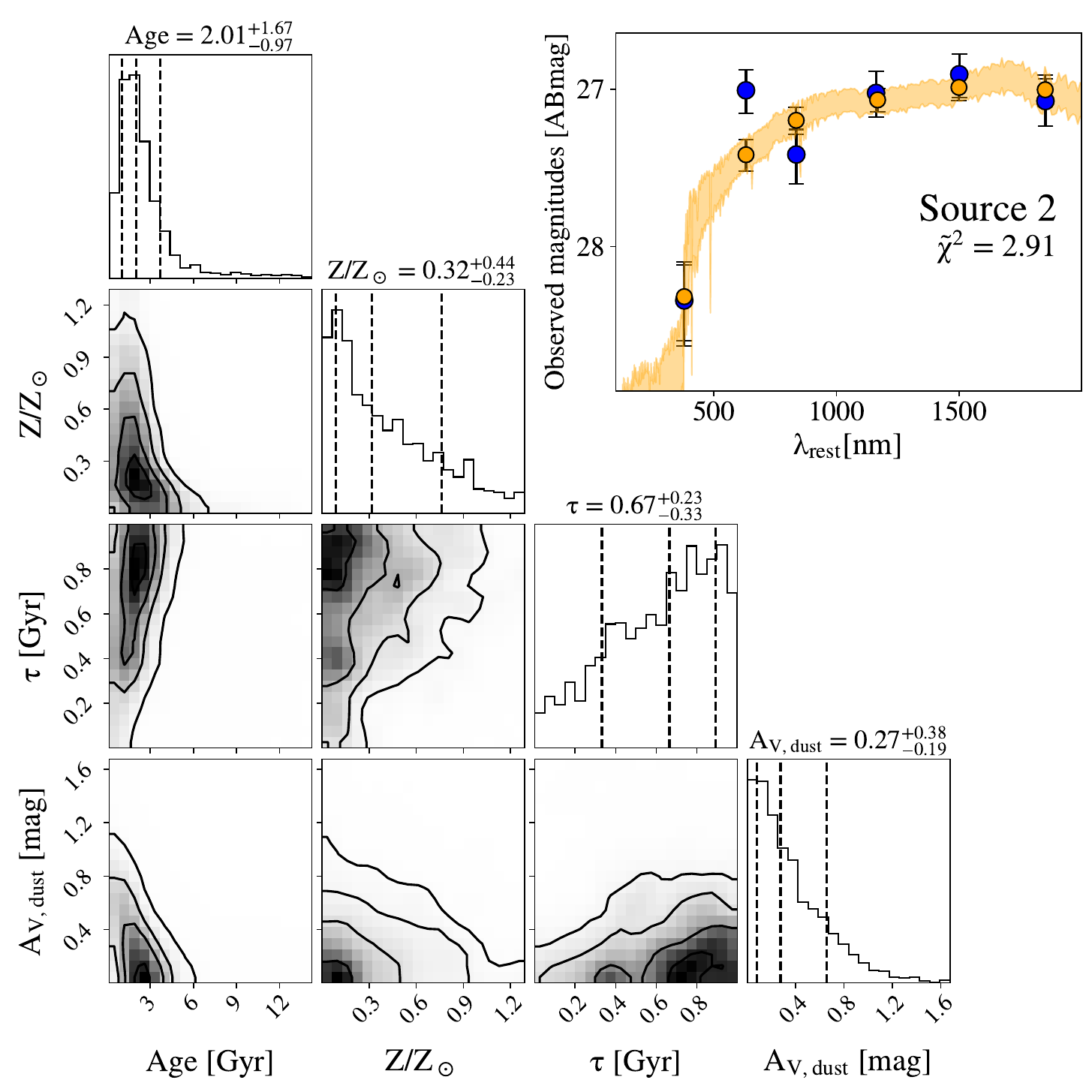} 
         \label{fig:2_corner}
    \end{subfigure}
    \begin{subfigure}
         \centering
         \includegraphics[width=0.4\textwidth]{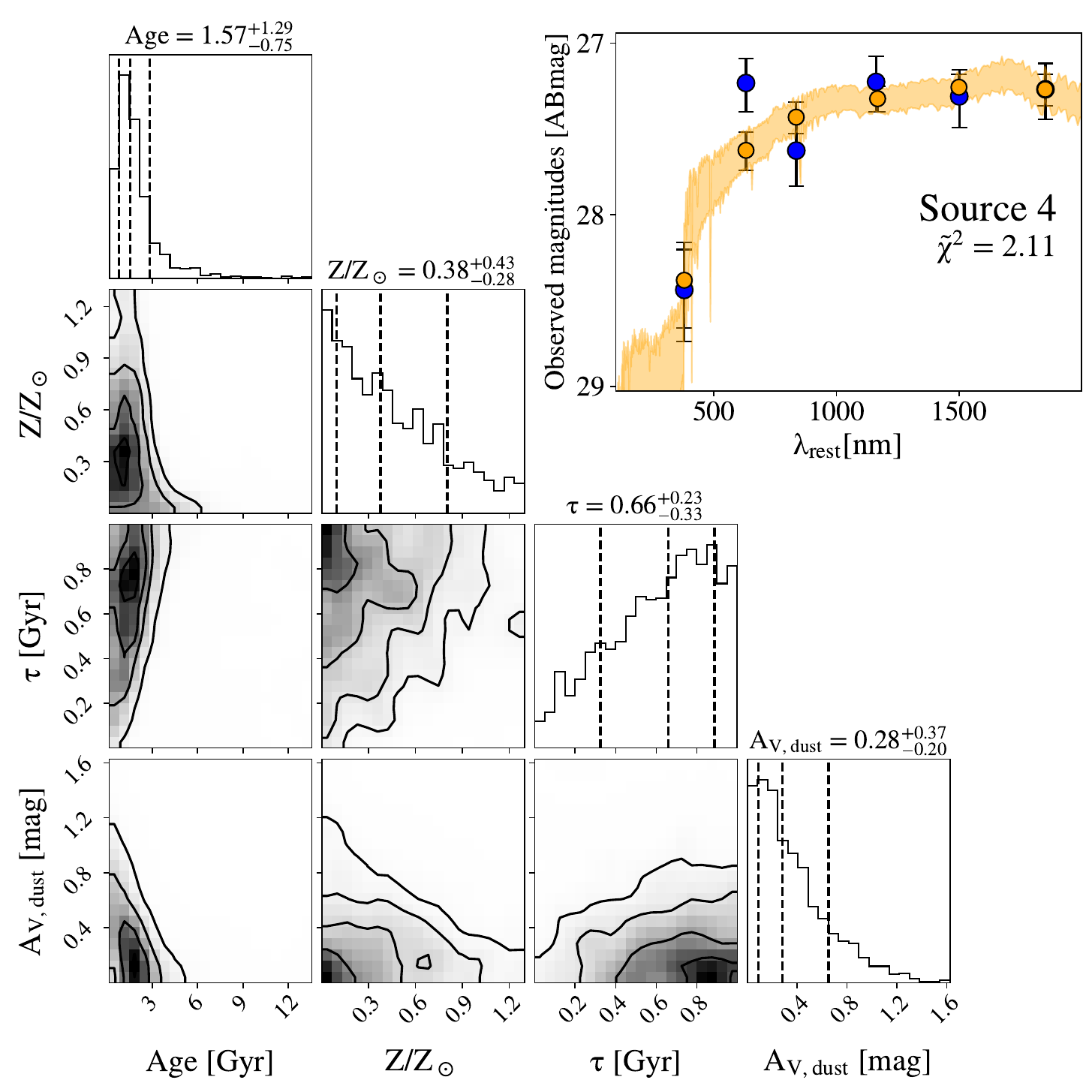} 
         \label{fig:4_corner}
    \end{subfigure}
    \begin{subfigure}
         \centering
         \includegraphics[width=0.4\textwidth]{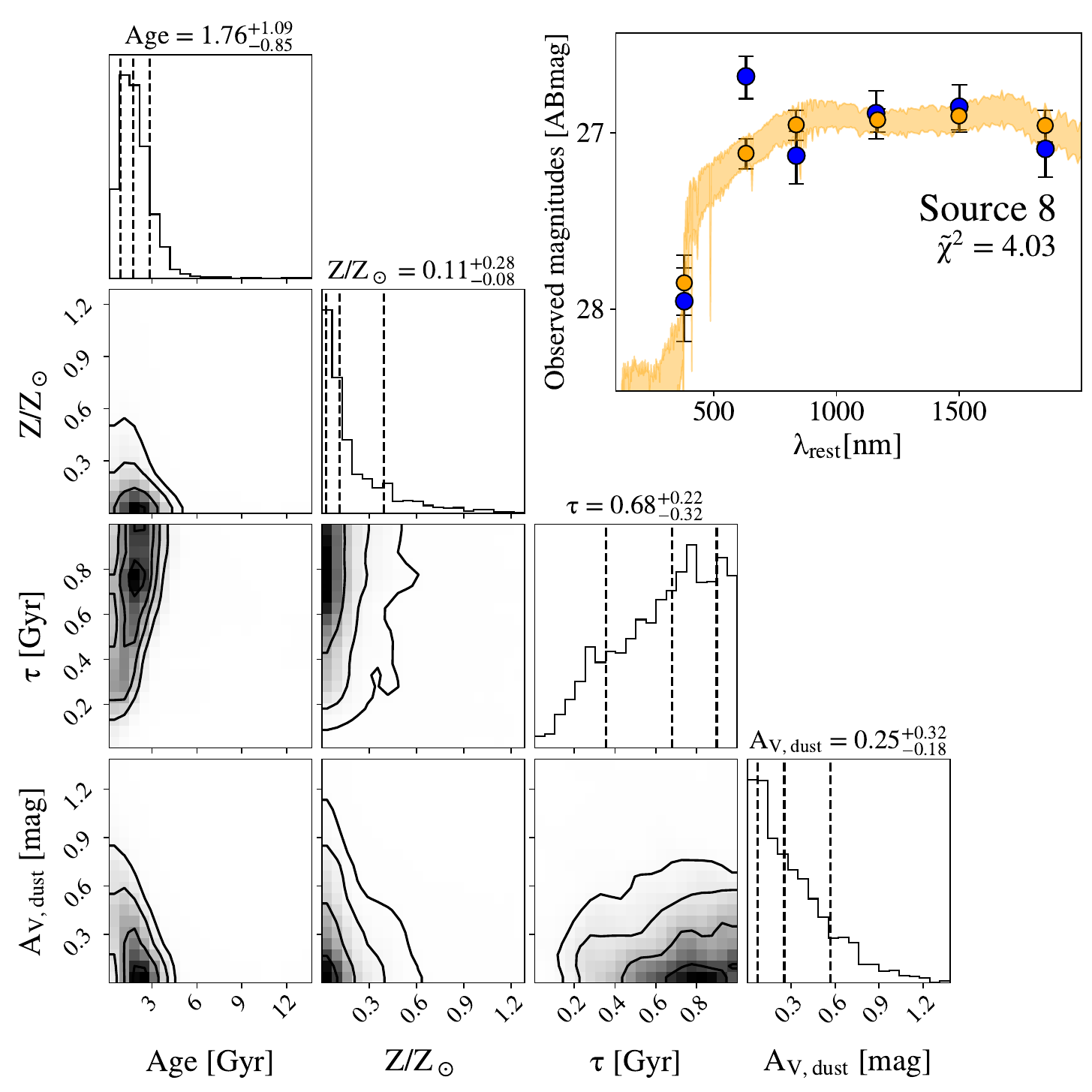} 
         \label{fig:8_corner}
    \end{subfigure}
    \begin{subfigure}
         \centering
         \includegraphics[width=0.4\textwidth]{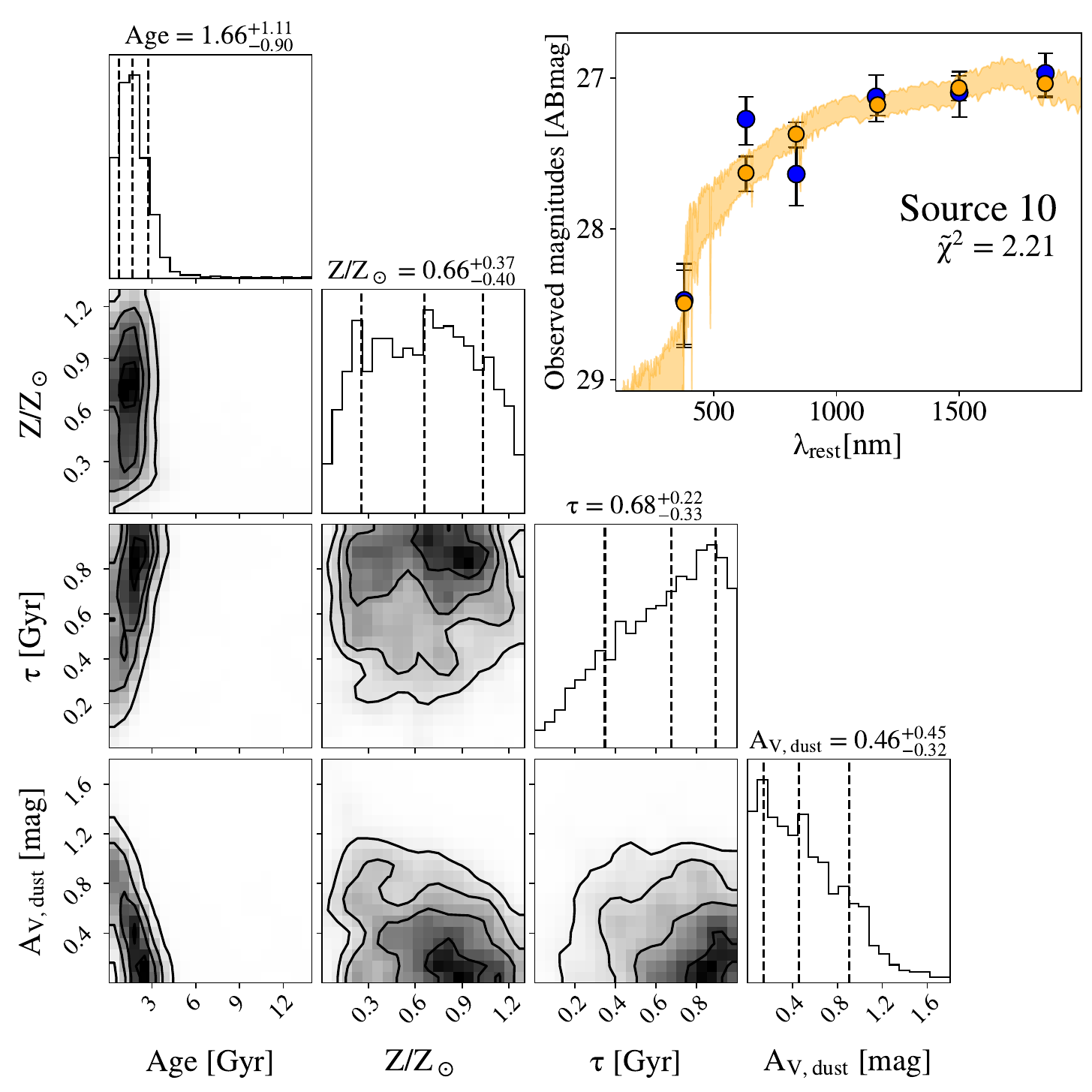} 
         \label{fig:10_corner}
    \end{subfigure}
    \label{fig:cornerplot}
    \caption{Joint posteriors for the four parameters studied for the Sparkler. The blue points are the data while the orange are the best fit and the corresponding spectrum from {\tt BAGPIPES}.}
\end{figure*}

To better illustrate the full parameter posteriors, we show in Fig.~\ref{fig:cornerplot} 2D contours for different parameter pairs. Note that the age is always very well constrained and that in general, all contours do close. The insets show the derived spectrum from \texttt{BAGPIPES} using the six photometric bands used in this study (blue points for the data and yellow points for the model).

All the best-fit spectra reproduce very well the observed photometry, except for the filter F150W, where the modeled flux is systematically underestimated. Thus, we tried redoing the analysis masking that photometric point, finding that ages are very stable, just $\sim$0.2 Gyr younger on average, and metallicities $\sim$0.2 higher.

\bibliographystyle{aa}
\bibliography{biblio}

%
%

\end{document}